# Millimeter-scale active area superconducting microstrip single-photon detector fabricated by ultraviolet photolithography

Guang-zhao Xu,[1,2,3] Wei-jun Zhang,[1,2,3*] Li-xing You,[1,2,3,4] Yu-ze Wang,[1,2,3] Jia-min Xiong,[1,2,3] Dong-Hui Fan,[1,2,3] Ling Wu,[1,3] Hui-qin Yu[1,3], Hao Li,[1,3] and Zhen Wang[1,3]

[1]*State Key Laboratory of Functional Materials for Informatics, Shanghai Institute of Microsystem and Information Technology, Chinese Academy of Sciences (CAS), Shanghai 200050, China*
[2]*Center of Materials Science and Optoelectronics Engineering, University of Chinese Academy of Sciences, Beijing 100049, China*
[3]*CAS Center for Excellence in Superconducting Electronics, Shanghai 200050, China*
[4]*State Key Laboratory of Marine Resource Utilization in South China Sea, Hainan University, No. 58, Renmin Avenue, Haikou, Hainan 570228, China*
*Corresponding author: zhangweijun@mail.sim.ac.cn*

**Abstract:** The effective and convenient detection of single photons via advanced detectors with a large active area is becoming significant for quantum and classical applications. This work demonstrates the fabrication of a superconducting microstrip single-photon detector (SMSPD) with a millimeter-scale active area via the use of ultraviolet (UV) photolithography. The performances of NbN SMSPDs with different active areas and strip widths are characterized. SMSPDs fabricated by UV photolithography and electron beam lithography with small active areas are also compared from the aspects of the switching current density and line edge roughness. Furthermore, an SMSPD with an active area of 1 mm × 1 mm is obtained via UV photolithography, and during operation at 0.85 K, it exhibits near-saturated internal detection efficiency at wavelengths up to 800 nm. At a wavelength of 1550 nm, the detector exhibits a system detection efficiency of ~5% (7%) and a timing jitter of 102 (144) ps, when illuminated with a light spot of ~18 (600) μm in diameter, respectively.

## 1. Introduction

Superconducting nanowire single-photon detectors (SNSPDs) can detect single photons in the ultraviolet (UV) to the mid-infrared range with excellent performance in terms of a high system detection efficiency (SDE) [1-3], low dark count rate (DCR) [4, 5], low timing jitter (TJ) [6, 7], and high counting rate (CR) [8, 9]. SNSPDs have been used in secure quantum communication [10], deep-space laser communication [11], depth imaging [12], and other fields. However, emerging application fields, like dark matter detection [13], confocal fluorescence imaging [14], and LIDAR [15], also increase the requirements for the large active area of the detectors. Conventional SNSPDs usually have an active area of 18 μm-in-diameter or less to match the single-mode fiber (SMF) alignment. The largest single pixel of an SNSPD is around 400 μm × 400 μm in size [16]. The large kinetic inductance prevents the further promotion of the active area of the single-pixel SNSPD [17]. Moreover, SNSPDs are generally fabricated using electron beam lithography (EBL), which is time-consuming and characterized by low throughput.

In recent years, with the expansion of studies on the dynamics of electrons and phonons in current-carrying superconducting strips, both theoretical investigations and experiments have shown that a superconducting microstrip with a strip width (usually 1-5 μm) smaller than the Pearl length can also detect single photons when the bias current is sufficiently close to the depairing current. This new type of detector is called a superconducting microstrip single-photon detector (SMSPD) [18-21]. Compared with the SNSPD, the SMSPD has a lower kinetic inductance and a lower requirement for the width precision of the strip, thereby enabling detector fabrication using photolithography. Previously, UV photolithography has been used to fabricate SNSPDs with strip widths 120-250 nm [22, 23]. Photolithography provides a new way to realize SMSPDs with an ultra-large active area and high throughput.

In 2020, Chiles et al. demonstrated the fabrication of a WSi SMSPD with saturated internal detection efficiency (IDE) at 1550 nm via UV i-line (365 nm) photolithography [24]. In 2021, Wollman et al. demonstrated several photolithography-fabricated WSi SMSPDs with a single-pixel active area of 3.1 mm × 3.1 mm and a strip width of 0.9 μm; however, the photon detection performance of these detectors has not yet been reported [25]. In 2022, Protte et al. demonstrated the fabrication of a WSi SMSPD via laser lithography; a device with a 1-μm-wide and 4-mm-long meandered strip exhibited saturated IDE at 1550 nm [26]. Most of the SMSPDs fabricated with photolithography employ WSi material, and the detection performance and yield of large-area SMSPDs fabricated via UV photolithography using other popular materials, such as NbN, remain unclear.

In this study, SMSPDs based on NbN material with an ultra-large active area of up to 1.5 mm × 1.5 mm are designed and fabricated based on UV photolithography. The detection performances of SMSPDs with different active areas and strip widths are characterized. Moreover, small-active-area SMSPDs fabricated by UV photolithography and EBL are compared in terms of the switching current density and line edge roughness (LER). The yield and detection performances of SMSPDs with an active area of 1 mm$^2$ are explored. At 0.85 K, illuminated with a laser spot of ~18 (600) μm in diameter, respectively, the SMSPD with an active area of 1 mm × 1 mm is found to exhibit a low TJ of 102 (144) ps and an SDE of ~5% (7%) at 1550 nm, as well as saturated IDE at wavelengths up to 800 nm.

## 2. Design and Fabrication

The wafer-stepper reticle was designed for the preliminary test of UV photolithography firstly. In the reticle, a total of 36 devices (called "small units") were arranged and included different device types, different strip widths, and different active area sizes. Specifically, the small units including three types of devices, namely microbridge, meandered, and spiral devices, have been described by Xu et al. [27] Each type of device contained strips with widths of 1, 2, and 3 μm, respectively. The meandered and spiral devices had different active areas with varied diameters of 0.1, 0.3, 0.5, 1, and 1.5 mm, respectively. To reduce the current-crowding effect in the bends [28], the meandered SMSPDs were designed with a filling factor of 0.33, and the spiral SMSPDs were designed with a filling factor of 0.5. Each device was tapered at the input and output pads, and a simple impedance-matching design was applied to the SMSPDs [29]. For comparison, the reduced reticle pattern was exposed by stepping at 12 different locations on a 4-inch silicon wafer. The "small unit" of these detectors was stepped repeatedly from one reticle. No obvious defect was observed at the stepper reticle.

In the experiment, superconducting NbN film with a thickness of 7 nm was first deposited on 4-inch double-sided thermal oxidized silicon wafers with a thickness of 400 μm via multi-chamber DC magnetron sputtering in a mixture of Ar and $N_2$ gases [30]. The target of the sputtering system was 8-inch Nb, and the film thickness variation on the 4-inch wafer was estimated to be less than 3.5%. After the NbN film deposition, a layer of UV photoresist (AZ703) was spin-coated on the surface of the NbN with a thickness of about 0.8 μm, and the NbN film was processed to form the designed patterns using an i-line photolithography stepper (FPA-3030i5+, Canon, ≤0.35 μm width resolution, write field of 0.5-22 mm) and reactive ion etching. Finally, the obtained wafer was diced, and the residual photoresist on the surface was removed to complete the fabrication. In contrast to EBL, the process of fabricating SMSPDs using a stepper is relatively simple and convenient. Only one step of exposure and etching is needed, i.e., the NbN active region and electrode pads are formed simultaneously, which greatly reduces the time consumption.

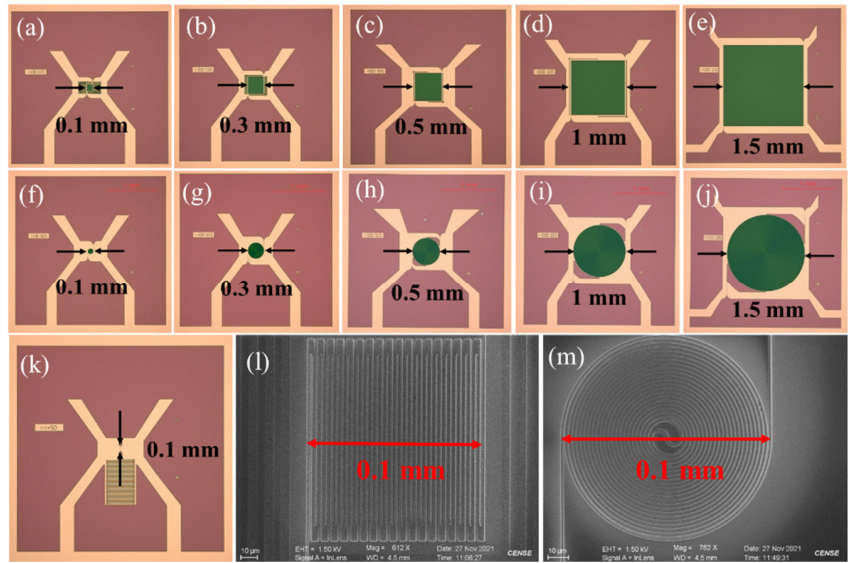

Figure 1. (a-k) Optical microscope images and (l, m) SEM images of the 1-μm-wide SMSPDs fabricated by the stepper. (a)-(e) Meandered devices with different active areas. (f)-(j) Spiral devices with different active areas. (k) A straight microbridge device. (l) A meandered device with a 100-μm side length. (m) A spiral device with a 100-μm diameter.

Figures 1(a)-1(k) display the typical optical microscope images of the different structures of the fabricated SMSPDs. Based on optical microscope inspection, there were no obvious dirty spots or pollution particles on the surface of the fabricated devices, and no obvious strip breaks were observed. The devices were further inspected with scanning electron microscopy (SEM). Figures 1(l) and 1(m) show the SEM images of the meandered and spiral SMSPDs with a diameter of 100 μm, which exhibited good overall strip width uniformity.

## 3. Performances Characterization of multiple-structure SMSPDs

First, the electrical properties of the devices in the two reduced reticle patterns distributed in the center of the wafer with different strip widths of 1, 2, and 3 μm were characterized at a temperature of 2.2 K. For the microbridge devices, the average switching currents of the devices with the three different strip widths were about 127.1±2.5, 258.3±3.8, and 390.2±5.2 μA, respectively. The critical temperature ($T_c$) of the microbridges was in the range of 7.3-7.9 K. As was expected, the switching current was found to be proportional to the strip width [31]. The switching currents of the meandered devices with different active areas and different strip widths on the same reduced reticle pattern were also characterized. As shown in Figure 2(a), as the active area increased (up to 1.5 mm × 1.5 mm), the switching current exhibited a downward trend. This is because the larger the area of the device, the more likely the appearance of defects in the long microstrip. Considering the average switching current of a 1-μm microbridge as a reference, as indicated by the dashed line in Figure 2(a), it was found to be basically the same as the microbridge switching current for SMSPDs with smaller active areas. The devices with 1- and 3-μm strip widths exhibited better switching current distribution uniformity in different active areas, while the devices with a 2-μm width achieved relatively poor performance. It should be noted that the measurement sampling points for the device shown in Figure 2(a) are small (2 samples per data point on average), and the devices located on two reticle patterns in the middle of the first wafers were only measured. Therefore, the measurement data in this part showed some randomness, which may result in some devices to exhibit high or low switching currents. For example, SMSPDs with active areas exceeding 1 mm$^2$ were also found to exhibit high switching currents comparable with those of the microbridges, especially at a 3-μm strip width. However, the 1.5 mm devices showed a relatively low switching current to the microbridge value. The reason may be attributed to the defects in the fabricated microstrips in these devices or the inhomogeneity of NbN film growth in the large 4-inch wafer. Furthermore, in devices with larger active areas (≥0.5 mm), meandered SMSPDs mostly outperformed spiral SMSPDs in terms of switching current (about 20% higher averagely) due to better width uniformity, as was observed by SEM. We speculated that the spiral patterns in UV exposure may cause some interference effects on the optical wavelength, which affects the quality of the strips. Therefore, subsequent studies on devices with ultra-large active areas mainly focused on meandered SMSPDs.

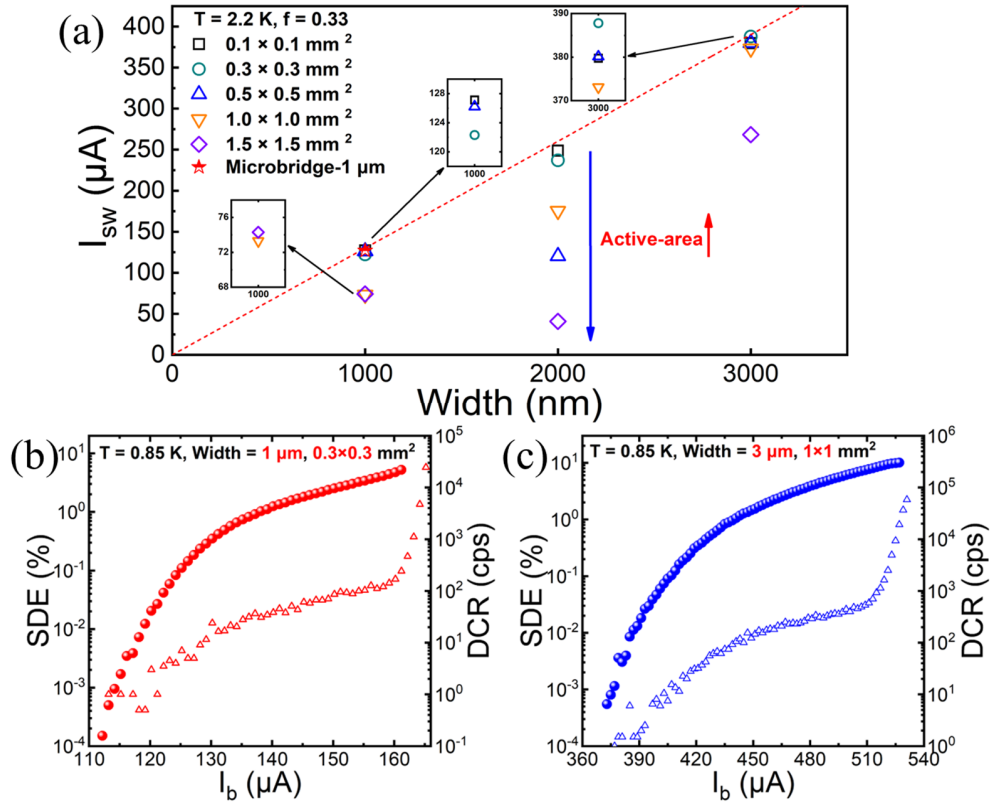

Figure 2. The characterization of different meandered SMSPDs fabricated by the stepper. (a) The switching current distributions of the meandered SMSPDs as a function of the strip width for different active areas. The dashed line is the average switching current distribution trend, which was deduced from the data of the 1-μm microbridges. The arrow indicates that the switching current decreased (blue) as the active area increased (red). Insets: zoomed-in areas of interest. Bias current-dependent SDE and DCR of the SMSPD illuminated at a 1550-nm wavelength: (b) with a strip width of 1 μm and an active area of 0.3 mm × 0.3 mm; (c) with a strip width of 3 μm and an active area of 1 mm × 1 mm.

SMSPDs with normal switching currents were first selected for preliminary testing. To achieve high switching currents and good single-photon detection sensitivity, experiments were performed on a compact closed-loop Gifford-McMahon (G-M) cryocooler at ~0.85 K, which was equipped with an adsorption refrigeration module. The method reported by Xu et al. [27] was adopted for the experimental setup. Especially, the device was connected in parallel with a shunt resistor (~5.8 Ω or ~2.0 Ω for 1-μm or 3-μm width devices, respectively) to prevent latching. The shunt resistor used in this study is smaller than that of that in Ref. [27], due to a larger switching current of the device. Note that we did not pursue the most suitable or optimal value for the shunt resistor in this study, which we estimate to be around ~1.9 Ω or ~0.7 Ω for 1-μm or 3-μm width devices, respectively. The shunt resistor was soldered onto a PCB with a distance of ~8 mm and packaged with the device in a small oxygen-free copper block. The device was connected with the golden pad on the PCB through wire bonding. The whole packaged block was mounted on the ~0.85 K cold plate. The electrical signal was transmitted through a cryogenic coaxial cable to a room-temperature amplifier using the SMA connectors.

Figure 2(b) demonstrates the detection performance of an SMSPD with a strip width of 1 μm and an active area of 0.3 mm × 0.3 mm. The device was capable of single-photon detection at a wavelength of 1550 nm. In the measurement of the first wafer, an SMF-28e fiber was used for optical coupling with a mode field diameter of ~10 μm in the whole experiment. The spot size of SMF-28e fiber here was enlarged to about 18 μm due to a long working distance of ~80 μm between the fiber facet and the device surface. Although the SDE as a function of the bias current was not saturated at this wavelength, the SDE of the SMSPD exceeded 5% with a DCR of about 100 cps. The detector can be operated over a wide range of bias currents. For comparison, the bias current-dependent SDE and DCR of the SMSPD with a strip width of 3 μm and an active area of 1 mm × 1 mm was also shown in Figure 2(c). It was observed that strip width had a relatively small impact on the photo-response performance of the SMSPDs. Regarding the detection performance of millimeter-scale devices, detailed measurements and analysis were performed in subsequent yield characterization batches. It is worth noting that the SMSPDs had not been treated with helium ion irradiation [32], so their IDE performance was relatively poor.

## 4. Comparison of the UV Photolithography and EBL Fabrication

To further evaluate the strip quality of SMSPDs fabricated by photolithography using a stepper, EBL with higher fabrication accuracy was used to tailor the wide microbridges processed by the stepper into thin microbridges in situ. Specifically, via the EBL overlay process (the overlay deviation was less than 40 nm), microbridges with smaller widths were exposed on the wide micron bridges, and were then developed and etched to obtain "in situ tailoring" devices. Figure 3(a) presents the principle of the "in situ tailoring" method, and Figure 3(b) shows an optical microscope image of the tailored microbridge. Via this method, the change of the switching current density before and after "in situ tailoring" on the same chip can be compared. The performance of microbridges with tailored microbridges in the same reticle pattern wafer with the same strip width could also be compared.

Figure 3(c) shows an SEM image of a tailored microbridge, which was tailored from having a strip width $w_{uvl}$ of 2.97 μm to having a narrower strip width $w_{ebl}$ of 0.99 μm. The widths before and after "in situ tailoring" are marked by white double arrows, respectively. The relatively bright areas in the long white double arrow in this figure are the remaining etch traces of the wider microbridge, corresponding to the line edge fabricated by UV photolithography. As shown in Figures 3(d) and 3(e), the LER data of the microbridges before and after "in situ tailoring" were extracted with ProSEM software via SEM segmental scanning with an accuracy of about 1-2 nm. LER is defined as the deviation of a feature edge from an ideal shape (e.g., a smooth and straight strip in our case). Mathematically, LER is defined as 3 times the standard deviation σ (LER = 3*σ) from the straight-line edge fit [33]. Thus, a 100-μm-long microbridge was divided into 24 segments. The ratios between the LER and strip width were calculated and plotted, and respectively correspond to the right axes in Figures 3(d) and 3(e). The average LER obtained by EBL $\delta_{ebl}$ was ~5.5 nm, and the average LER obtained by UV photolithography $\delta_{uvl}$ was ~14.3 nm (i.e., about 2.5 times the value of $\delta_{ebl}$). It was observed that if the microstrips are fabricated by the same method as EBL or UV photolithography, the corresponding absolute LER will be approximately independent of the strip width. Moreover, the ratios of LER/$w_{ebl}$ and LER/$w_{uvl}$ were distributed in the range of 0.4%-0.6%, and LER/$w_{ebl}$ was slightly larger than LER/$w_{uvl}$.

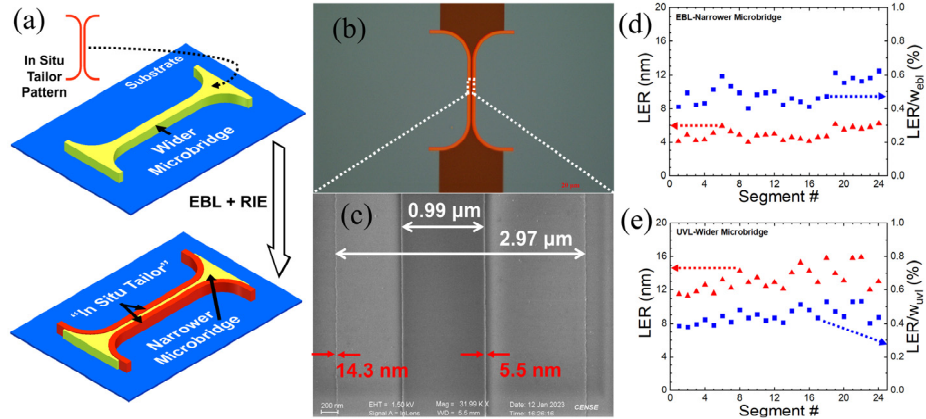

Figure 3. (a) A schematic diagram of the "in situ tailoring" method, via which a stepper-fabricated wide microbridge is tailored into a narrow microbridge by the EBL and reactive ion etching (RIE) processes. (b) An optical microscope image of a 100-μm-long microbridge after "in situ tailoring". (c) An enlarged SEM image of a segment of the tailored microbridge, for which the strip width was reduced from ~3 μm to ~1 μm, as marked with white double arrows, and the average LER values are indicated with red arrows. (d) The LER (red triangles) and LER/$w_{ebl}$ (blue squares) of the narrower microbridge tailored by EBL, as obtained from 24 segments. (e) The LER (red triangles) and LER/$w_{uvl}$ (blue squares) values of the wider microbridge fabricated by UV photolithography, as obtained from the corresponding segments.

Measured at 2.2 K, the switching current of the wide microbridge before tailoring was about 390.40 μA, and that of the narrow microbridge after tailoring was about 129.28 μA. The switching current densities of the wide and narrow microbridges were 18.77 and 18.66 mA/μm$^2$, respectively, with a difference of about 0.59%. Furthermore, the switching current of the tailored narrow microbridge was within the range of the averaged switching current (~127.1 ± 2.5 μA) for the 1-μm-wide microstrips described in Section 3. In addition, the average switching currents of the microstrips with different strip width in Section 3 were fitted by a linear function, then a residual strip width of 34.72 ± 3.27 nm was observed in the intercept of the x-axis, which represented around 17.4 nm on each edge of the microstrip even when the switching current got to zero. Ideally, the fitting line should pass through the origin. This residual strip width is close to the independently measured value of ~14.3 nm LER in Figure 3(e). This result proves that our method of analyzing switching current from LER aspect is reasonable and reliable. In terms of processing damage effects, for microscale-width structures, we found that there was no significant difference in film square resistance at 20 K $R_{sq}$(20 K) ~650 Ω/sq and $T_c$ ~7.7 K before and after the "in situ tailoring" processing. In the future, a more thorough comparison might have included an EBL and UV photolithography bridge of the same thickness fabricated next to each other. These experimental results indicate that the wide microstrips of the small-active-area SMSPDs obtained from EBL and UV photolithography had comparable uniformity.

## 5. Yield and Performances Characterization of Millimeter-scale SMSPDs

Production yield is a key factor in the achievement of large-scale manufacturing. To evaluate the yield of the devices fabricated using UV photolithography, a new reticle pattern was designed for the devices with millimeter-scale active areas. Figure 4(a) shows a test die in the designed reticle pattern. Sixteen of these identical test dies were arranged in a single reticle. Each test die contained three independent devices, namely a 1-mm$^2$ meandered microstrip arranged in the middle and two reference microbridges with a 3-μm-wide, 100-μm-long strip distributed on the left and right sides. For the SMSPD with a 1-mm$^2$ active area, the strip width and filling factor were 3 μm and 0.33, respectively. Figure 4(b) shows the overall device layout of a 4-inch silicon wafer. The reduced reticle patterns were exposed at 12 different square parts (parts 1-12) on the wafer. There was a total of 192 devices distributed across the wafer. The four central parts (parts 1-4) of the wafer contained 64 large-active-area SMSPDs and 128 reference microbridges, which were used to characterize the yield. During actual use, devices in the wafer center will be more likely to be selected for application experiments. The $T_c$ value of the reference microbridges was in the range of 7.6-8.1 K. The switching current distribution of the devices, including the reference microbridges in two central top parts (parts 1-2) and the SMSPDs with the millimeter-scale active area in parts 1-4, were then characterized at the temperature of 2.2 K, as shown in Figures 4(c) and 4(d), respectively. According to the statistics of the switching current without a shunt resistor, the average switching current of the microbridges was generally at a relatively high value of about 508.2 μA, which is close to a Gaussian distribution, with a standard deviation of about 45.1 μA (a ratio of ~8.9%). The difference in switching current of two wafers was mainly due to the deposition of $Nb_xN_{1-x}$ films with different stoichiometric ratios since the average critical temperature of the two films differed slightly (<0.3 K) and the measurement temperature was the same (both at 2.2 K). We estimated the microbridge yield to be 67.2% (43/64) based on the criterion that its switching current exceeds the average of the switching currents of the all measured microbridges on the second wafer. However, the switching current of the SMSPDs with a 1-mm$^2$ active area was at a relatively low level and had a wide distribution range. The switching currents of these SMSPDs were all below 480 μA.

Figure 4(e) shows the planar spatial distribution of the SMSPDs with a 1-mm$^2$ active area for parts 1-4 in the center of the wafer. No obvious position-dependent regularity was observed for the positional distribution of the switching currents. Therefore, according to the results exhibited in Figures 4(c)-4(e), the switching current reduction of the SMSPDs with a 1-mm$^2$ active area may have been caused by some randomly distributed defects in the NbN film. We have confirmed through SEM inspections that millimeter-scale devices with relatively small switching currents did have some obvious defects. These defects are not easily captured by the microbridges, but by large-area devices.

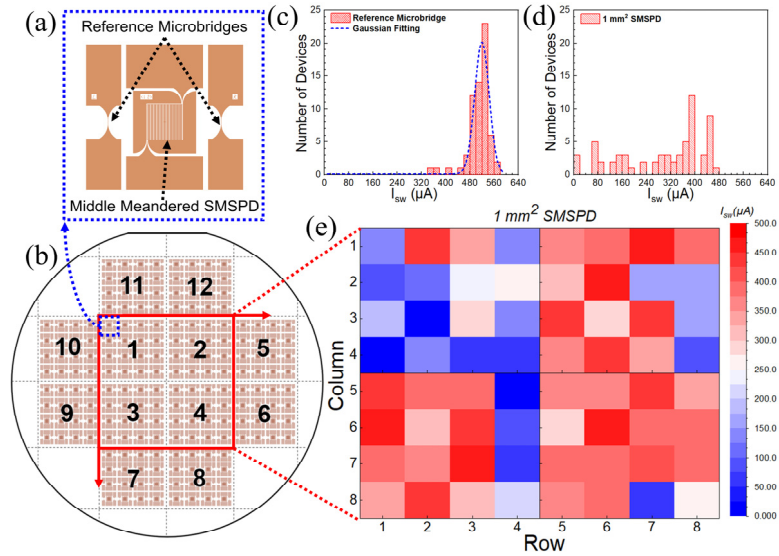

Figure 4. The yield characterization of SMSPDs with a 1-mm$^2$ active area. (a) A test die layout containing two reference microbridges and a middle meandered SMSPD with a 1-mm$^2$ active area. (b) The overall device layout for a reduced reticle pattern array on a 4-inch silicon substrate, which consisted of 12 numbered parts. (c) The switching current distribution of the reference microbridges for parts 1-2. (d) The switching current distribution of the 1-mm$^2$ SMSPDs in central parts 1-4. (e) The planar spatial distribution of the 1-mm$^2$ SMSPDs in parts 1-4. Redder squares represent devices with a larger switching current, and bluer squares represent devices with a smaller switching current.

One SMSPD with a 1-mm$^2$ active area and a switching current of ~467 μA at 2.2 K was selected for further characterization. The depairing current $I_{dep}(T = 2.2\ K) = \frac{w[1.76k_bT_c]^{3/2}}{eR_{sq}\sqrt{\hbar D}}[1-(\frac{T}{T_c})^2]^{3/2} = 684$ μA of this SMSPD was determined according to the reported methods [18], by using the physical parameters measured in our experiments: $D$ = 0.48 cm$^2$/s, $T_c$ = 7.8 K, $R_{sq}$(20 K) = 656 Ω/sq. The ratio of $I_{sw}/I_{dep}$ of the SMSPD under test was found to be 0.68.

The device was then tested in a 0.85-K system and shunted with a ~2.0-Ω resistor. Figure 5(a) shows the current-voltage (I-V) curve of the device with a switching current of about 610 μA, which is much higher than the one of un-shunted. We attributed this phenomenon to two reasons: firstly, the un-shunted device would indeed show a lower measured switching current than its true value due to latching; secondly, after the shunt resistor was connected, the nominal switching current of the device would change, that is, the display value will increase, which is caused by the current shunting effect [19, 27]. In addition, the shunted current can be influenced by factors such as a contact resistance in the wire bonds, and resistance in the cables, requiring further study.

Figure 5(b) demonstrates the SDE performance of the SMSPD coupled with an SMF-28e fiber with different light spot sizes. By increasing the working distance between the fiber facet and the device surface by 80 μm or 3 mm, the diameter ($Φ$) of the light spot illuminated on the device was increased to ~18 μm or ~600 μm (floodlight illumination), respectively. The laser spot was focused on the middle position of the 1 mm active area, and no obvious shift was observed before and after the thermal cycle. The SDE as a function of the bias current was recorded at two different input photon fluxes of 0.1 M and 1 M photons per second, respectively. When illuminated with $Φ$ ~18(~600)-μm light spot, the SDEs under the two-photon fluxes were respectively about 5% (7%) and 3% (4%) at 610 μA with a DCR of about 300 cps. The reduction of the SDE under the high input photon flux could be due to the blocking effect caused by the slow response of the detector. The low SDE can be owing to the low IDE and low optical absorption efficiency. If a cavity design based on distributed Bragg reflector is adopted [34], the calculated absorption efficiency at 1550 nm wavelength could be improved to ~60% for the 3-μm-wide SMSPD with a 0.33 filling factor through the numerical simulation of COMSOL software.

Figure 5(c) shows the comparisons of the DCR behaviors when the device was coupled with different light spot sizes and without the fiber, respectively. Here the intrinsic DCR was measured by removing the coupling fiber and shielding the chip-mounted block with aluminum tape. As shown by the square red dots in Figure 5(c), the intrinsic DCR of the SMSPD was over 15 times lower than those measured with fiber at the bias current of <0.9$I_{sw}$. Note that small differences in the DCR curves when coupled with different fiber spot sizes were caused by slightly different intensities of ambient stray light during the measurement. However, we noticed that the measured intrinsic DCR does not exhibit the expected monotonic logarithm linearity across the bias current region. We speculated that the intrinsic DCR deviation with $I_b$ smaller than 0.93$I_{sw}$ may be due to the influence of the electrical noise of our measurement system or the vortex movement modulated by the shunt resistance. This issue needs to be further confirmed in future study.

Figure 5(d) shows the current dependence of the normalized detection efficiency (DE) measured at different incident wavelengths. Illuminated with the $Φ$ ~18-μm light spot, the logistic function fitting reveals that near-saturated detection was achieved at an 800-nm wavelength with an estimated IDE of 99.8%, and an estimated IDE of 97.5% (98.5% for $Φ$ ~600-μm light spot) was achieved at an 850-nm wavelength. At 532 nm, the normalized single-photon response curve fluctuated slightly at a high bias current due to the instability of the light source. Although there was no obvious saturation plateau at the wavelengths of 1150 and 1350 nm, the estimated IDE could also exceed 60% and 20%, respectively, when illuminated with the $Φ$ ~18-μm light spot.

Figure 5(e) shows the normalized histogram count distribution of the time delay between the pulsed laser and signal of the SMSPD at 1550 nm, and TJ is regarded as the full width at half-maximum (FWHM) of normalized counts. Notably, the Gauss fitting result reveals a TJ of about 102 ps and 144 ps when illuminated with ~18 μm and ~600 μm light spot, respectively, at a high bias current of 610 μA. A single Gaussian peak appeared in the histogram, thus indicating the good single-photon response of the SMSPD. The TJ increase with the light spot size was possible due to the increase in geometrical TJ.

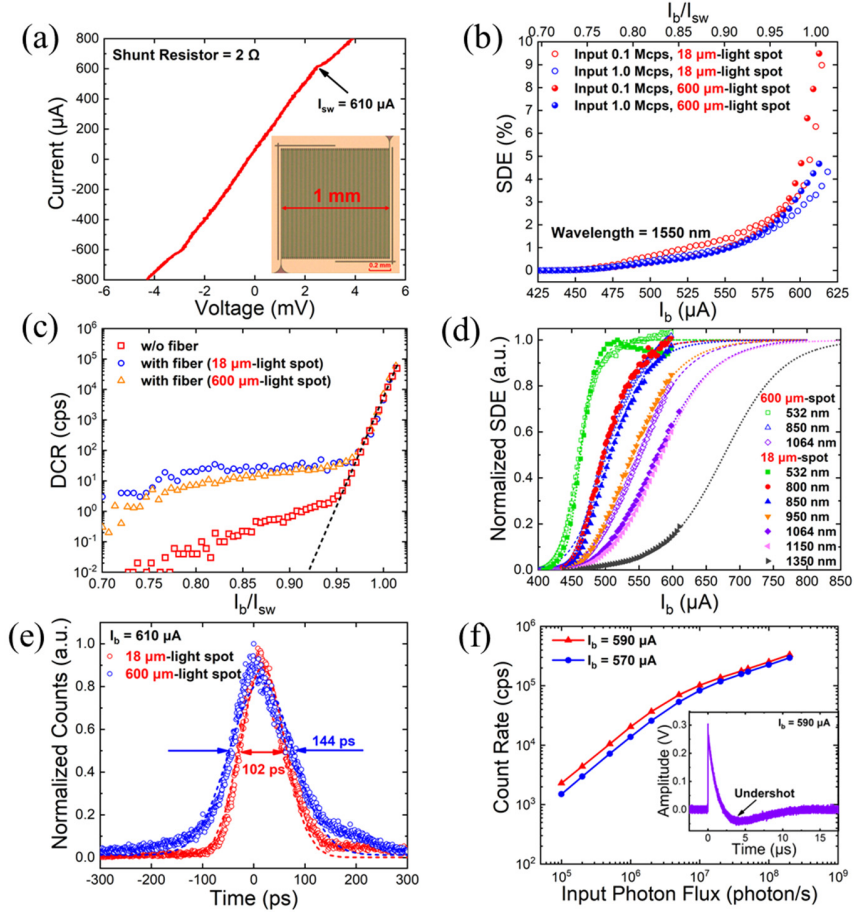

Figure 5. Detection performance characterization of the 1-mm$^2$ active area SMSPD, illuminated with a light spot of $\Phi$ ~18 μm and 600 μm, respectively. (a) I-V curve measured with a shunt resistor. The inset is a microscope image of the active area of the device. (b) SDE versus the bias current at different input photon fluxes at 1550 nm. (c) DCR of the SMSPD coupled with fiber (different light spot sizes) and without fiber (i.e., the intrinsic DCR) as a function of the normalized bias current ($I_b/I_{sw}$). The dashed line is used to guide the eye. (d) Normalized DE versus the bias current at different wavelengths. (e) The histogram of time-correlated single photon counts measured at 1550 nm and biased at 610 μA, with an FWHM of 102 ps and 144 ps, respectively. (f) Count rate of the device as a function of the input photon flux biased at 590 and 570 μA, respectively, illuminated with the 18 μm light spot. Inset: response pulse waveform. An undershoot appears on the falling edge.

Figure 5(f) shows the count rate dependence of the input photon flux measured at bias currents of 590 and 570 μA, and a maximum count rate of about 0.33 Mcps was obtained at the maximum incident photons (around $2\times10^8$ photons/s) before the occurrence of output response oscillations. The relatively low maximum count rate could be attribution to two possible reasons: first is the bias current-dependent unsaturated SDE, resulting in that the detector has to wait more than one decay time for the SDE to recover. Second, recent studies [35] implied that the limited bandwidth of the RF amplifier in the readout circuit can affect the device's pulse recovery time and lead to an undershoot in the output pulse, causing the device to latch earlier. Inset of Figure 5(f) showed a significant undershot in the output pulse of our device recorded at a bias current of 590 μA [36]. A pulse magnitude of 305 mV was gained with a good signal-to-noise ratio of the output signal. The detection performances prove that this kind of large-active-area SMSPD could satisfy some application requirements in the UV and visible wavelengths, such as satellite laser ranging and dark matter detection.

Table 1. Comparison of performances of large-area SMSPDs operated at 1550 nm wavelength

| Materials | Lithography | Area (μm$^2$) | Width (μm) | SDE (%) | TJ (ps) | Yield (%) |
|---|---|---|---|---|---|---|
| MoSi$_x$[20] | EBL | 400×400 | 1 | ≪6 | N/A | N/A |
| WSi[24] | EBL (UV lithography for microbridge) | 362×362 | 2 | N/A | N/A | N/A |
| WSi[26] | Laser lithography | 120×120 | 1 | N/A | 225 | N/A |
| NbN (This paper) | UV lithography | 1000×1000 | 3 | ~5 | 102 | 39.1 |

The presented SMSPD exhibited saturated IDE at a 532-nm wavelength with a bias current of about 500 μA, corresponding to a ratio of 0.82 for the switching current with a shunt resistor. As a rough estimation, the yield of these 1-mm$^2$-active-area SMSPDs was defined as having a switching current greater than 0.82 times that of the presented SMSPD at 2.2 K. In other words, an SMSPD with a switching current exceeding 383 μA is expected to saturate at 532 nm. Thus, the yield of the SMSPDs was estimated to be 39.1% (25/64). The results of the yield measurement of the SMSPDs demonstrate that it is feasible to fabricate SMSPDs with millimeter-scale active areas by UV photolithography with NbN thin film, although the current yield and IDE are still restricted by factors such as the NbN film quality. In Table 1, the performances of the large-active-area SMSPDs at 1550 nm wavelength in recent researches are compared. The characterization of single-photon response abilities and yield for the large-active-area SMSPD in this study are displayed comprehensively.

In the future, the IDE and yield of the device will be further improved via multiple approaches, such as ion irradiation [32], chemical composition adjustment (e.g., conversion of δ-NbN into γ-Nb$_4$N$_3$ [37]), and deposition parameter optimization. Meanwhile, the thin film quality will be improved with less-substrate-dependent material (e.g., NbTiN) [2]. The authors are optimistic about the realization of a millimeter-scale SMSPD with saturated single-photon detection at 1550 nm via the use of UV photolithography.

## 6. Conclusions

A large-active-area NbN SMSPD based on UV photolithography with an i-line stepper was designed, fabricated, and characterized. SMSPDs were fabricated with different active areas and different strip widths on a 4-inch silicon substrate. It was found that the larger the SMSPD active area, the more difficult it is to achieve the switching current of the microbridge with the same width, which is limited by the quality of the NbN thin film. Compared with the SMSPD fabricated by EBL, the differences in the switching current density and LER between the two methods were correspondingly characterized. As a demonstration, the electrical properties of the SMSPDs with a 1-mm$^2$ active area (64 devices in total) were systematically characterized. Furthermore, illuminated with a light spot of ~18 (600) μm in diameter, this type of device exhibited an SDE of ~5% (7%) at a wavelength of 1550 nm, a TJ of 102 (144) ps, respectively, and a near-saturated IDE up to an 800-nm wavelength, thus presenting a yield of 39.1% as determined from the switching current criterion for the visible wavelength. Therefore, compared with EBL, UV photolithography significantly improves the time cost and process flow of SMSPD fabrication. This technology is promising in the study and development of ultra-large-active-area SMSPDs with high throughput.

**Funding**

This work was supported by the National Natural Science Foundation of China (NSFC, Grant No. 61971409), the National Key R&D Program of China (Grant No. 2017YFA0304000), and the Science and Technology Commission of Shanghai Municipality (Grant Nos. 18511110202 and 2019SHZDZX01). W. -J. Zhang is supported by the Youth Innovation Promotion Association, CAS (Grant No. 2019238).

**Acknowledgments**

The authors would like to thank Xiaoyu Liu and Jia Huang for technical assistance with EBL, Yu Wu for information about the NbN thin film roughness, Zhimin Guo for proofreading the article, Liliang Ying for technical assistance with UV photolithography, and Photon Technology (Zhejiang) Co. for testing environment.

**Disclosures**

The authors declare that they have no conflicts of interest.

**Data and materials availability**

All data are available in the article.

**References**

1. P. Hu, H. Li, L. You, H. Wang, Y. Xiao, J. Huang, X. Yang, W. Zhang, Z. Wang, and X. Xie, "Detecting single infrared photons toward optimal system detection efficiency," Opt. Express **28**, 36884-36891 (2020).
2. J. Chang, J. W. N. Los, J. O. Tenorio-Pearl, N. Noordzij, R. Gourgues, A. Guardiani, J. R. Zichi, S. F. Pereira, H. P. Urbach, V. Zwiller, S. N. Dorenbos, and I. Esmaeil Zadeh, "Detecting telecom single photons with 99.5−2.07+0.5% system detection efficiency and high time resolution," APL Photonics **6**, 036114 (2021).
3. D. V. Reddy, R. R. Nerem, S. W. Nam, R. P. Mirin, and V. B. Verma, "Superconducting nanowire single-photon detectors with 98% system detection efficiency at 1550 nm," Optica **7**, 1649-1653 (2020).
4. H. Shibata, K. Shimizu, H. Takesue, and Y. Tokura, "Ultimate low system dark-count rate for superconducting nanowire single-photon detector," Opt. Lett. **40**, 3428-3431 (2015).
5. W. J. Zhang, X. Y. Yang, H. Li, L. X. You, C. L. Lv, L. Zhang, C. J. Zhang, X. Y. Liu, Z. Wang, and X. M. Xie, "Fiber-coupled superconducting nanowire single-photon detectors integrated with a bandpass filter on the fiber end-face," Supercond. Sci. Tech. **31**, 035012 (2018).
6. B. Korzh, Q.-Y. Zhao, J. P. Allmaras, S. Frasca, T. M. Autry, E. A. Bersin, A. D. Beyer, R. M. Briggs, B. Bumble, M. Colangelo, G. M. Crouch, A. E. Dane, T. Gerrits, A. E. Lita, F. Marsili, G. Moody, C. Peña, E. Ramirez, J. D. Rezac, N. Sinclair, M. J. Stevens, A. E. Velasco, V. B. Verma, E. E. Wollman, S. Xie, D. Zhu, P. D. Hale, M. Spiropulu, K. L. Silverman, R. P. Mirin, S. W. Nam, A. G. Kozorezov, M. D. Shaw, and K. K. Berggren, "Demonstration of sub-3 ps temporal resolution with a superconducting nanowire single-photon detector," Nat. Photonics **14**, 250-255 (2020).
7. J. P. Allmaras, A. G. Kozorezov, B. A. Korzh, K. K. Berggren, and M. D. Shaw, "Intrinsic Timing Jitter and Latency in Superconducting Nanowire Single-photon Detectors," Phys. Rev. Appl. **11**, 034062 (2019).
8. W. Zhang, J. Huang, C. Zhang, L. You, C. Lv, L. Zhang, H. Li, Z. Wang, and X. Xie, "A 16-Pixel Interleaved Superconducting Nanowire Single-Photon Detector Array With A Maximum Count Rate Exceeding 1.5 GHz," IEEE Trans. Appl. Supercond. **29**, 2200204 (2019).
9. M. Perrenoud, M. Caloz, E. Amri, C. Autebert, C. Schönenberger, H. Zbinden, and F. Bussières, "Operation of parallel SNSPDs at high detection rates," Supercond. Sci. Tech. **34**, 024002 (2021).
10. K. J. Wei, W. Li, H. Tan, Y. Li, H. Min, W. J. Zhang, H. Li, L. X. You, Z. Wang, X. Jiang, T. Y. Chen, S. K. Liao, C. Z. Peng, F. H. Xu, and J. W. Pan, "High-Speed Measurement-Device-Independent Quantum Key Distribution with Integrated Silicon Photonics," Phys Rev X **10**, 031030 (2020).
11. L. You, J. Quan, Y. Wang, Y. Ma, X. Yang, Y. Liu, H. Li, J. Li, J. Wang, J. Liang, Z. Wang, and X. Xie, "Superconducting nanowire single photon detection system for space applications," Opt Express **26**, 2965-2971 (2018).

12. L. Kong, Q. Zhao, K. Zheng, H. Lu, S. Chen, X. Tao, H. Wang, H. Hao, C. Wan, X. Tu, L. Zhang, X. Jia, L. Kang, J. Chen, and P. Wu, "Noise-tolerant single-photon imaging with a superconducting nanowire camera," Opt Lett **45**, 6732-6735 (2020).
13. Y. Hochberg, I. Charaev, S. W. Nam, V. Verma, M. Colangelo, and K. K. Berggren, "Detecting Sub-GeV Dark Matter with Superconducting Nanowires," Phys. Rev. Lett. **123**, 151802 (2019).
14. F. Xia, M. Gevers, A. Fognini, A. T. Mok, B. Li, N. Akbari, I. E. Zadeh, J. Qin-Dregely, and C. Xu, "Short-Wave Infrared Confocal Fluorescence Imaging of Deep Mouse Brain with a Superconducting Nanowire Single-Photon Detector," ACS Photonics **8**, 2800-2810 (2021).
15. Y. Guan, H. Li, L. Xue, R. Yin, L. Zhang, H. Wang, G. Zhu, L. Kang, J. Chen, and P. Wu, "Lidar with superconducting nanowire single-photon detectors: Recent advances and developments," Opt. Las. in Engineer. **156**, 107102 (2022).
16. J. Chiles, I. Charaev, R. Lasenby, M. Baryakhtar, J. Huang, A. Roshko, G. Burton, M. Colangelo, K. Van Tilburg, A. Arvanitaki, S. W. Nam, and K. K. Berggren, "New Constraints on Dark Photon Dark Matter with Superconducting Nanowire Detectors in an Optical Haloscope," Phys Rev Lett **128**, 231802 (2022).
17. S. Steinhauer, S. Gyger, and V. Zwiller, "Progress on large-scale superconducting nanowire single-photon detectors," Appl Phys Lett **118**, 100501 (2021).
18. D. Y. Vodolazov, "Single-Photon Detection by a Dirty Current-Carrying Superconducting Strip Based on the Kinetic-Equation Approach," Phys. Rev. Appl. **7**, 034014 (2017).
19. Y. P. Korneeva, D. Y. Vodolazov, A. V. Semenov, I. N. Florya, N. Simonov, E. Baeva, A. A. Korneev, G. N. Goltsman, and T. M. Klapwijk, "Optical Single-Photon Detection in Micrometer-Scale NbN Bridges," Phys. Rev. Appl. **9**, 064037 (2018).
20. I. Charaev, Y. Morimoto, A. Dane, A. Agarwal, M. Colangelo, and K. K. Berggren, "Large-area microwire MoSi single-photon detectors at 1550 nm wavelength," Appl Phys Lett **116**, 242603 (2020).
21. M. Ejrnaes, C. Cirillo, D. Salvoni, F. Chianese, C. Bruscino, P. Ercolano, A. Cassinese, C. Attanasio, G. P. Pepe, and L. Parlato, "Single photon detection in NbRe superconducting microstrips," Applied Physics Letters **121**, 262601 (2022).
22. A. D. Beyer, S. W. Nam, W. H. Farr, M. D. Shaw, F. Marsili, M. S. Allman, A. E. Lita, V. B. Verma, G. V. Resta, J. A. Stern, and R. P. Mirin, "Tungsten Silicide Superconducting Nanowire Single-Photon Test Structures Fabricated Using Optical Lithography," IEEE Trans. Appl. Supercond. **25**, 2200805 (2015).
23. J. M. Shainline, S. M. Buckley, N. Nader, C. M. Gentry, K. C. Cossel, J. W. Cleary, M. Popovic, N. R. Newbury, S. W. Nam, and R. P. Mirin, "Room-temperature-deposited dielectrics and superconductors for integrated photonics," Opt Express **25**, 10322-10334 (2017).
24. J. Chiles, S. M. Buckley, A. Lita, V. B. Verma, J. Allmaras, B. Korzh, M. D. Shaw, J. M. Shainline, R. P. Mirin, and S. W. Nam, "Superconducting microwire detectors based on WSi with single-photon sensitivity in the near-infrared," Appl Phys Lett **116**, 242602 (2020).
25. E. E. Wollman, V. B. Verma, A. B. Walter, J. Chiles, B. Korzh, J. P. Allmaras, Y. Zhai, A. E. Lita, A. N. McCaughan, E. Schmidt, S. Frasca, R. P. Mirin, S. W. Nam, and M. D. Shawa, "Recent advances in superconducting nanowire single-photon detector technology for exoplanet transit spectroscopy in the mid-infrared," J. Astron. Telesc. Instrum. Syst. **7**, 011004 (2021).
26. M. Protte, V. B. Verma, J. P. Hopker, R. P. Mirin, S. W. Nam, and T. J. Bartley, "Laser-lithographically written micron-wide superconducting nanowire single-photon detectors," Supercond. Sci. Tech. **35**, 055005 (2022).
27. G.-Z. Xu, W.-J. Zhang, L.-X. You, J.-M. Xiong, X.-Q. Sun, H. Huang, X. Ou, Y.-M. Pan, C.-L. Lv, H. Li, Z. Wang, and X.-M. Xie, "Superconducting microstrip single-photon detector with system detection efficiency over 90% at 1550 nm," Photonics Res. **9**, 958-967 (2021).
28. I. Charaev, A. Semenov, S. Doerner, G. Gomard, K. Ilin, and M. Siegel, "Current dependence of the hot-spot response spectrum of superconducting single-photon detectors with different layouts," Supercond. Sci. Tech. **30**, 025016 (2017).
29. D. Zhu, M. Colangelo, C. Chen, B. A. Korzh, F. N. C. Wong, M. D. Shaw, and K. K. Berggren, "Resolving Photon Numbers Using a Superconducting Nanowire with Impedance-Matching Taper," Nano Lett **20**, 3858-3863 (2020).
30. W. Zhang, L. You, H. Li, J. Huang, C. Lv, L. Zhang, X. Liu, J. Wu, Z. Wang, and X. Xie, "NbN superconducting nanowire single photon detector with efficiency over 90% at 1550 nm wavelength operational at compact cryocooler temperature," Sci. China Phys. Mech. Astron. **60**, 120314 (2017).
31. Y. P. Korneeva, N. N. Manova, M. A. Dryazgov, N. O. Simonov, P. I. Zolotov, and A. A. Korneev, "Influence of sheet resistance and strip width on the detection efficiency saturation in micron-wide superconducting strips and large-area meanders," Supercond. Sci. Tech. **34**, 084001 (2021).
32. W. Zhang, Q. Jia, L. You, X. Ou, H. Huang, L. Zhang, H. Li, Z. Wang, and X. Xie, "Saturating Intrinsic Detection Efficiency of Superconducting Nanowire Single-Photon Detectors via Defect Engineering," Phys. Rev. Appl. **12**, 044040 (2019).
33. G. P. Patsis, V. Constantoudis, A. Tserepi, E. Gogolides, and G. Grozev, "Quantification of line-edge roughness of photoresists. I. A comparison between off-line and on-line analysis of top-down scanning electron microscopy images," J. of Vac. Sci. & Tech. B **21**, 1008-1018 (2003).
34. L. You, H. Li, W. Zhang, X. Yang, L. Zhang, S. Chen, H. Zhou, Z. Wang, and X. Xie, "Superconducting nanowire single-photon detector on dielectric optical films for visible and near infrared wavelengths," Supercond. Sci. Tech. **30**, 084008 (2017).

35. J. S. Luskin, E. Schmidt, B. Korzh, A. D. Beyer, B. Bumble, J. P. Allmaras, A. B. Walter, E. E. Wollman, L. Narváez, V. B. Verma, S. W. Nam, I. Charaev, M. Colangelo, K. K. Berggren, C. Peña, M. Spiropulu, M. Garcia-Sciveres, S. Derenzo, and M. D. Shaw, "Large active-area superconducting microwire detector array with single-photon sensitivity in the near-infrared," arxiv, 2303.10739 (2023).
36. H. Hao, Q.-Y. Zhao, L.-D. Kong, S. Chen, H. Wang, Y.-H. Huang, J.-W. Guo, C. Wan, H. Liu, X.-C. Tu, L.-B. Zhang, X.-Q. Jia, J. Chen, L. Kang, C. Li, T. Chen, G.-X. Cao, and P.-H. Wu, "Improved pulse discrimination for a superconducting series nanowire detector by applying a digital matched filter," Appl. Phy. Lett. **119**, 232601 (2021).
37. Y. Pan, H. Zhou, X. Zhang, H. Yu, L. Zhang, M. Si, H. Li, L. You, and Z. Wang, "Mid-infrared $Nb_4N_3$-based superconducting nanowire single photon detectors for wavelengths up to 10 μm," Opt. Express **30**, 40044-40052 (2022).